\documentclass{eas}
\usepackage{graphicx}
\begin{document}

\title{Cosmic Rays in the Disk and Halo of Galaxies}
\author{V.A.Dogiel}\address{P.N. Lebedev Institute, Leninskii
pr, 53, 119991 Moscow, Russia}
\author{D. Breitschwerdt }\address{Zentrum f\"ur Astronomie und Astrophysik, Technische Universit\"at Berlin, Hardenbergstr. 36, D-10623 Berlin, Germany }

\runningtitle{Cosmic Rays in the Disk and Halo of Galaxies}
\begin{abstract}
We give a review of cosmic ray propagation models. It is shown
that the development of the theory of cosmic ray origin leads
inevitably to the conclusion that cosmic ray propagation in the
Galaxy is determined by effective particle scattering, which is
described by spatial diffusion. The Galactic Disk is
surrounded by an extended halo, in which cosmic rays are confined
before escaping into intergalactic space. For a long time cosmic
ray convective outflow from the Galaxy (galactic wind) was believed to be
insignificant.  However, investigations of hydrodynamic
stability and an analysis of ISM dynamics (including cosmic rays)
showed that a galactic wind was emanating near the disk, and accelerating towards the
halo, reaching its maximum velocity far away from the disk.
Therefore convective cosmic ray
transport should be important in galactic halos. Recent analysis
of the gamma-ray emissivity in the Galactic disk of EGRET data,
which showed that cosmic rays are more or less uniformly
distributed in the radial direction of the disk, as well as the interpretation of
soft X-ray emission in galactic halos, give convincing evidence of the existence of a
galactic wind in star forming galaxies.
\end{abstract}
\maketitle

\section{Historical Review}
The story of cosmic rays (CRs) begins about 1900 when it was found that
electroscopes (which measured the amount of ionization) discharged,
even if they are kept in the dark well away from sources of
natural radioactivity (for the history of CRs see the excellent
monographs of \cite{long}). In order to elucidate the role of the
Earth, Hess and Kohlh\"{o}rster carried out manned balloon flights.
Especially successful was the flight of August 7, 1912 (birthday
of CRs) when V. Hess reached an altitude of 5 km. He found that the
ionization rate there was several times higher than observed at
sea level. Hess wrote:
\begin{quote}
The results of the present observations seem to be mostly explained
by the assumption that a radiation of
very high penetrating power enters our atmosphere from above.
\end{quote}
This radiation was  named by Millikan in 1925 as {\it cosmic
radiation} or {\it cosmic rays (CRs)}.

The first characteristics of CRs were derived from measurements of
their flux near Earth. It is quite amazing, how many global
parameters of CR propagation in the Galaxy may be inferred from
these "local" measurements. The most striking characteristic was
that the spectrum of CRs in the energy range, covering six orders
of magnitude, was found to be a power-law with a single index of
$\gamma\simeq -2.7$, thereby strongly constraining the possible CR
acceleration mechanisms.
Another very interesting result obtained from measurements was
that a group of light elements like Li, Be, B was enhanced by
about five orders of magnitude in comparison with e.g. solar
abundances. Although the fraction of these light elements in the total
flux of CRs is negligibly small, we shall see how much they allow us
to learn about CR propagation in the Galaxy.

The first question arising from the early measurements was: where are CRs
produced, in our galaxy, or outside? In models of extragalactic CR origin, particles
observed near Earth traveled from extragalactic sources to our
Galaxy before reaching the solar system. Then, at least in the
vicinity of the Galaxy (or, even, in the local group of galaxies)
the  density of CRs should be the same as near Earth, i.e.
$w_{cr}\sim 10^{-12}$ erg cm$^{-3}$. It is then easy
to predict the expected flux of gamma-rays from $\pi^\circ$ decays
from the nearest galaxies, i.e., the Small and Large magellanic Cloud (SMC, LMC),
since the total mass of hydrogen there is known. In 1972 Ginzburg
presented this criterion and estimated the flux from LMC for the
extragalactic model, which was about $10^{-7}$ ph
cm$^{-2}\,$s$^{-1}\,$ for $E_\gamma>100$ MeV. Only in 1993 an upper
limit of the gamma-ray flux was obtained for this galaxy (\cite{sree}),
which was about $<0.5\cdot 10^{-7}$ ph cm$^{-2}\,$s$^{-1}$. This
means that the CR density in the local extragalactic medium is
nonuniform, and CRs (at least below $10^{15}$ eV as it turns out)
are produced in the Galaxy.
As potential Galactic sources supernova (SN) explosions were assumed.
This idea was suggested by Baade and Zwicky in 1934. Then Ginzburg and Syrovatskii (1964)
substantiated it in their classical monograph \emph{``The origin of CRs''}.

\section{ The Leaky Box Model}
The interpretation of observational data required a simple model of CR
evolution, which was put forward and subsequently called leaky-box or thick target model (for a
review see e.g. Cesarsky, 1980). In this model
effects of particle spatial propagation were neglected, meaning
that either the particle spatial distribution was uniform or the evolution
of particle spectra was "local" when the spatial coordinates are
parameters only.
Particles were thought to be trapped within a box with
semipermeable walls, at which they were reflected, but also having
a finite probability to escape ("leak") into intergalactic space.
In the stationary case the
equation for the CR density $N$ with energies $E$ has a
very simple form
\begin{equation}
\frac{d}{dE}\left(\frac{dE}{dt}N\right)+\frac{N}{T}=Q(E) \,.
\end{equation}
Here $Q$ describes particle injection (source term), $dE/dt$ stands for
energy losses, and $T$ is the characteristic time of catastrophic
losses or particle escape.

These equations allowed to interpret the above-mentioned
overabundance of the nuclei from the Li, Be, B group. The
over-abundance of light elements, which are not produced in the
interior of stars, was thought to be caused by fragmentation in
the ISM by inelastic collisions of CR primaries with background
particles.  Then the ``source term'' for  secondary nuclei of sort
$i$ by primary nuclei of sort $j$ can be written as
$Q_i(E)=\sum\limits_j \bar{n}_Hv\sigma_{ij}N_j$, where $n_H$ is
the average density of the background gas, $v$ is the velocity of
primary nuclei, and $\sigma$ is the cross-section of
fragmentation. Since continuous energy losses are negligible for
relativistic CRs the density ratio of secondary $N_s$ to primary
nuclei $N_p$ is
\begin{equation}
\frac{N_s}{N_{pr}}\simeq \bar{n}_H\sigma c T \,.
\end{equation}
The value $x=\bar{\rho}cT$, where $\bar{\rho}=\bar{n}_Hm_p$, $m_p$
is the proton rest mass, is  denoted as the grammage. From the
observed ratio ${N_s}/{N_{pr}}$ near Earth it was derived that
$x\simeq 10$ gr cm$^{-2}$ (see, e.g. \cite{fer}).

With this value of $x$, and, what is quite noteworthy, from
measurements of the CR flux \emph{near Earth}, it is possible to
derive the total luminosity of CRs in the Galaxy. It is
$L_{cr}=W_{cr}/T$, where $W_{cr}$ is the total energy of CRs in
the Galaxy, and $T$ is the average lifetime of CRs, both values
being unknown. Let us multiply the numerator and the denominator
by $\bar{n}_H m_p c$ (where $m_p$ is the proton rest mass), and
write $W_{cr}$ as $\bar{w}_{cr}V$, where $w_{cr}$ is the average
energy density of CRs in the Galaxy and $V$ is the volume of the
Galaxy filled with CRs. 
Then the product $M_H=\bar{n}_H V m_p$ is the
total mass of hydrogen in the Galaxy, which is known from radio
measurements. Assuming that the density of CRs does not
vary much over the volume of Galaxy, the total CR luminosity becomes
$L_{cr}=w_{cr} M_H c/x$.
All values of this ratio are known from observations: $w_{cr}\sim
10^{-12}$ erg cm$^{-3}$, $M_H\simeq 10^{43}$ g, and $x\simeq 10$
gr cm$^{-2}$. Then the CR luminosity is an astonishing
$L_{cr}=3\cdot10^{40}~\mbox{erg s$^{-1}$}$, amounting to about 3\% of
the total Galactic SN (hydrodynamic) energy release rate!

Potential sources of CRs in the Galaxy are observed to have the following
luminosities: 1. SNe: $10^{42}$ erg s$^{-1}$; 2. neutron
stars: $10^{41}$ erg s$^{-1}$; 3. stellar winds of O/B stars:
$10^{41}$ erg s$^{-1}$; 4. flare stars: $3\cdot 10^{40}$ erg
s$^{-1}$, i.e. only about several percents of the energy of SNe
should be transformed into the flux of CRs in order to provide
their luminosity in the Galaxy.

Some isotopes of the Li, Be, B group are radioactive. For instance the
isotope $^{10}Be$ decays into $^{10}B$ within the characteristic half life
time $\tau_r=2.2\cdot 10^6$ years. If the decay time is of the
order of the CR lifetime $T$, then the latter can be estimated from
the density ratio of stable secondary nuclei $N^s$ to radioactive
secondary nuclei $N^r$. Indeed, from the leaky-box model this ratio
is
\begin{equation}
\frac{N_{l}^r}{N_{i}^s}=\frac{1/T_l}{1/T_l+1/\tau_r}\frac{\sum\limits_jn_Hv
\sigma_{lj}^rN_j^p}{\sum\limits_jn_Hv\sigma_{ij}^sN_j^p} \,.
\label{tl}
\end{equation}
From measurements it was derived that the value of CR
lifetime calculated in the framework of the leaky-box model is $T_l\sim
2\div 3\cdot 10^7$ years (see e.g. Ahlen et al. 2000).
Two very important conclusions follow from the values of
$x$ and $T$:
\begin{itemize}
\item In $10^7$ years CRs cover a distance
$10^{25}$cm, or 3 Mpc. On the other hand, the thickness of the
galactic disk is only about 300-500 pc and its radius is about
10-15 kpc. Then, the trajectories of CRs in the Galaxy should be
strongly tangled, i.e they perform a random walk with many
scatterings off magnetic field irregularities. Their propagation
can simply be described by a  diffusion in coordinate space;
\item If we estimate the average density of the gas from the
values of $x$ and $T$ we obtain the value $\bar{n}_H\simeq 0.25$
cm$^{-3}$, which is almost four times smaller than the average gas density
in he galactic disk. This means that CRs spend most of
their lifetime outside the galactic disk, in the so-called
galactic halo surrounding the disk where the gas density is small.
Since the half thickness of the disk is about 250 pc, the
thickness of the halo should be at least 1 kpc.
\end{itemize}
Thus, it was concluded that the description of CRs in the
framework of the leaky-box model is strongly restricted, and it
was necessary to abandon the leaky-box model.

\section{The Diffusion Model}
The idea about particle scattering in the interstellar medium
(ISM) by magnetic fields was suggested by Fermi (1949), being a
necessary condition for stochastic acceleration of CRs in the ISM.
In 1964 Parker investigated qualitatively the process of charged
particle scattering by magnetic fluctuations. He analyzed particle
propagation along the magnetic field lines whose equations was:
$y^\ast=C_1+F(x)$, $z^\ast =C_2$, where $y^\ast$ and  $z^\ast$ are
coordinates of  magnetic field lines and the constant $C_1$ and
$C2$ specify the line. 
Thus, we have a uniform magnetic field, ${\bf e}_xB_0$, in the
x-direction and a magnetic field irregularity in the y-direction,
${\bf b}_y={\bf e}_y dF(x)/dx$. The equation for propagation of a
particle with the charge $Ze$, the mass $M$ and the velocity $\bf
v$ is
\begin{equation}
\frac{d{\bf v}}{dt}= \frac{Ze}{Mc} \nabla \times \left[{\bf v}\cdot{\bf
B}\right]
\end{equation}
 Below one should use magnitudes:
 ${\bf B}={\bf B}_0+{\bf b}_y$, where $|{\bf b}| \ll |{\bf B}_0|$.
Then the particle displacement in y-direction is
\begin{equation}
y=\Omega\int\limits_{-\infty}^t d\tau F(v_\|\tau)\sin(\Omega\tau) \,,
\end{equation}
where $\Omega=ZeB_0/Mc$ and $v_\|$ is the velocity component along
$x$-direction. For the irregularity in the form
$F(x)=a\exp(-x^2/b^2)$ a displacement of the particle pitch angle
(angle between  vectors of magnetic field and particle momentum)
is derived from
\begin{equation}
\frac{\triangle
v_\bot}{v_\|}=\sqrt{\pi}\frac{a}{b}\frac{b^2}{R^2}\exp\left(-\frac{b^2}{4R^2}\right) \,,
\label{vrat}
\end{equation}
where $R=v/\Omega$ is the particle Larmor radius. One can see from
this equation that interactions of particles with magnetic
perturbations are resonant, and particles are scattered most
efficiently by irregularities when the scale of magnetic field fluctuations
is of the order of the particle Larmor radius, $a\sim R$.

We conclude from Eq.~(\ref{vrat}) that a single interaction of a
particle with a resonant irregularity changes its pitch angle
$\theta$ by $\triangle \theta\sim b/B_0$. Since particles are
scattered by fluctuations randomly the total angle displacement
after $N_{sc}$ interactions is proportional to
$\sqrt{N_{sc}} b/B_0$,  and therefore in order to be scattered
through the angle $\pi$, we have the condition $\sqrt{N_{sc}}
b/B_0 \sim \pi$.
The distance necessary to be scattered through
$\pi$ is $\lambda_{sc} \sim N_{sc} R \sim (B_0/b)^2R$. Then we can
introduce a diffusion coefficient for this random walk of a
particle along magnetic field lines $D \sim \lambda_{sc} v$.

In 1965 Parker published a paper, in which he presented a
phenomenological diffusion equation for CRs in interplanetary
space. He concluded "that the feature of magnetic field, which
determines the nature of the propagation of energetic particles is
the presence of small-scale irregularities. The irregularities
appear with dimensions comparable to the radius of gyration. The
irregularities scatter, or reflect, the energetic particles back
and forth along the line of force of the large-scale field, so
that there is no tendency for the particles to move systematically
in either direction. Then the effect of magnetic irregularities is
to cause the CR particles to random walk. The random walk of the
CR particles is a Markov process, describable by a Fokker-Planck
equation:"
\begin{equation}
\frac{\partial N}{\partial t}+\frac{\partial }{\partial
x_j}(Nv_j)+\frac{\partial }{\partial E}\left(\frac{d E}{d
t}N\right)-\frac{\partial }{\partial
x_j}\left(K_{ij}\frac{\partial N}{\partial x_i}\right)=0 \,.
\label{eqd}
\end{equation}
A formal and correct derivation of the diffusion approximation for
multiple scattering of charged particles in a random magnetic
field was performed by Dolginov and Toptygin in 1966 (see
translation of this paper \cite{dolg}). Unfortunately, this
analysis is almost unknown in the astrophysical community.

In the ISM magnetic irregularities are plasma
waves whose magnetic field strength can be described as a
spectrum
\begin{equation}
{\bf b}(t,{\bf r})=\sum\limits_\alpha\int
d^3k\exp\left[-i\omega^\alpha({\bf k})t+i{\bf k}\cdot({\bf
r})\right]{\bf b}^\alpha({\bf k}) \,,
\end{equation}
where $\alpha$ denotes a certain species of plasma waves, which
is determined from the corresponding dispersion relation
$\omega=\omega^\alpha({\bf k})$.
This relation together with the condition of resonant
particle-wave interaction, which in this case has the form
\begin{equation}
\omega^\alpha({\bf k})-k_\|v_\|\mp\omega_H=0 \,,
\label{res}
\end{equation}
determines the conditions of particle scattering in the ISM.
Here $\omega_H=\Omega/\gamma$ is the relativistic gyrofrequency.
The frequency of particle scattering (over pitch angles) is in this
case
\begin{equation}
\nu_\mu^\alpha\simeq2\pi^2\mid\omega_H\mid\frac{k_{res}W_{res}^\alpha(k_{res})}{B_0^2} \,.
\end{equation}
The value of $k_{res}$ is determined from the
resonant condition (\ref{res}), which for the case of MHD waves has
the form $k_{res}=1/(R\mid\mu\mid)$. Here $\mu$ is the cosine of particle pitch angle, and
$W(k)=b^2(k)/4\pi$ is the wave energy density.

If scattering is so effective that the flux of propagating
particles is izotropized  then the Fokker-Plack equation can be
simplified and reduced to the diffusion equation. The diffusion
coefficient along the magnetic field line is calculated from the
equation
\begin{equation}
D=v^2\left[\frac{9}{2}\int\limits_0^1d\mu(1-\mu^2)\nu_\mu\right]^{-1} \,;
\end{equation}
This approximation can definitely be used for cosmic rays because
their flux, as follows from observation, is highly isotropized,
for details of the diffusion approximation, see Kulsrud and Pearce
(1969) or the monograph by Berezinskii et al. (1990).

The major advantage of using the diffusion equation to describe
particle propagation is the possibility to properly analyze the
various cosmic ray components whose distribution in the Galaxy is
strongly nonuniform. As an example, we mention the result of
Prishchep and Ptuskin (1975) who showed that $T_l$ derived from
the leaky-box model, Eq.(\ref{tl}), underestimated strongly the CR
lifetime. What we infer from this model is a combination of the
real CR lifetime $T_D$ (which in the framework of the diffusion
model is $T_D\sim h^2/D$, where $h$ is a size of the halo) and the
radioactive decay time $\tau_r$, $T_l\sim \sqrt{T_D\tau_r}$ .

The diffusion approximation allowed to analyze the parameters of
nonthermal emission generated by different CR components
in different regions of the galactic disk and halo. In this
way it was possible to estimate the size of the CR halo $h$. The
diffusion model successfully described the measured parameters of the CR
flux near Earth and the spectral characteristics of radio and
gamma-ray emission (for details, see the monograph of Berezinskii et
al. (1990), and for a similar, but more detailed, analysis based on the
numerical program GALPROP of Strong and collaborators in the
review of Strong et al., 2007)). However, this analysis rested
on two phenomenological simplifications whose validity was not
proven: a) it was supposed that all CRs independently of
their energy and charge escape into the extragalactic space from
the same surface of the halo, and, b) the effect of CR convective
transport was ignored. Below we present arguments that these
simplifications are in general not appropriate.

\section{Boundary Conditions on the Halo Surface}
In the  monographs of Ginzburg and Syrovatskii (1964) and
Berezinskii et al. (1990) (as well as in the GALPROP program) it
was assumed that the boundary conditions for all components of
CRs at the halo surface $\Sigma$ is $N\mid_\Sigma=0$, i.e.
all CR particles escape from the same surface. Dogiel
et al. (1993, 1994) tried to derive these boundary conditions from
kinetic equations. They analyzed the spherically symmetric case for the
magnetic field, which is described by the equation:
\begin{equation}
\mu v\frac{\partial f}{\partial
r}+v\frac{1-\mu^2}{r}\frac{\partial f}{\partial
\mu}=\frac{\partial }{\partial
\mu}\left[(1-\mu^2)\nu(r,E,\mu)\frac{\partial f}{\partial
\mu}\right]+Q(E)\delta(r) \,. \label{sph}
\end{equation}
By variable transformation this equation can be reduced to the
case of particle propagation along a divergent magnetic flux
tube.

The source of particles, $Q$, is assumed to be at the coordinate
origin, and $\nu(r,E)$ is the scattering frequency. The problem
was analyzed for an infinite space and it was supposed that the
frequency of particle scattering decreases with the distance from
the sources. Then the solution can be expanded in a Legendre
polynomial series: $f(r,E,\mu)=f_0(r,E)+ \delta^{-1}
f_1(r,E,\mu)+\delta^{-2} f_2(r,E,\mu)+...$, where
$\delta=\nu/c(\partial f/\partial r)$ . It was shown that for
$\delta \gg 1$ the kinetic equation reduces to the diffusion
equation with the coefficient $D\sim c^2/3 \nu$. However, on the
surface, where $\delta\sim 1$, scattering is not sufficient to
isotropize the distribution function. For the frequency
$\nu=\nu_0(E/E_0)^{-\alpha}(r/r_0)^{-b}$ this surface is located
at:
\begin{equation}
\bar{r}(E)\simeq
r_0\left[\frac{\nu_0(E/E_0)^{-\alpha}r_0}{c}\right] \label{rh} \,.
\end{equation}
At distances larger than $\bar{r}(E)$ the solution of Eq. (\ref{sph}) is strongly
anisotropic
\begin{eqnarray}
&&f(r,E,\mu)=\frac{C(E)\xi}{\xi+\tau(r,E)}
\exp\left[-\frac{(1-\mu)(r/\bar{r})^2}{\xi+\tau(r,E)}\right],\nonumber\\
&&\tau(r,E)=\frac{4\nu_0(E/E_0)^{-\alpha}(r/\bar{r})^{3-b}}{3-b}\,,
\end{eqnarray}
and the parameters $C$ and $\xi$ are determined from the conditions
at $r=\bar{r}$.

The conclusion of this analysis is that CRs escape freely
(i.e. run-away particles) from this surface along the magnetic
field lines with their proper velocities.

Moreover, if magnetic fluctuations in the halo are excited by
CRs due to the streaming instability whose increment is (Cesarsky 1980)
\begin{equation}
\Gamma(k)=\frac{\Omega}{\gamma}\frac{N(>\bar{E})}{n}\left(\frac{\bar{V}}{V_A}-1\right) \,,
\label{stinst}
\end{equation}
where $\bar{E}$ is derived from the resonant condition, $n$ is the
density of the background plasma, $\bar{V}$ is the average velocity of
the CR flux long the magnetic field lines, and
$V_A=B_0/\sqrt{4\pi n m_p}$ is the Alfv\'en velocity), the
position of the run-away surface is determined by a single parameter, i.e.
by the power of CR sources: the higher the power, the further away
from the sources is the  run-away (halo) surface. It was also shown
that the halo size is a function of CR energy, and thus the
halo radius decreases when the particle energy increases.

\section{Cosmic Ray Propagation: The Effect of Convection ("Galactic
Wind")}

For a long time the effect of convective transport of CRs (the
second term in Eq.(\ref{eqd})) was not considered as significant,
though in a number of papers the effect of convection on the CR
spectrum was analyzed (see Bulanov et al. 1972, Jokipii 1976,
Dogiel et al. 1980 etc.). The reasons were the following: a) the
diffusion model was generally successful in accounting for most CR
observations, both direct observations of particles near Earth as
well as indirect observations like the diffuse radio-synchrotron
and gamma-ray emission of our  Galaxy; b) investigations did not
imply that convective transport could be completely excluded, but
wind velocities in the Galactic disk were only of the order of 10 km
s$^{-1}$ or less (for comparison the escape velocity from the Galaxy is
about 500 km s$^{-1}$).

However, a series of publications drew more attention to the
process of CR transport (galactic wind). The first one was
a paper by Wentzel (1974), who analyzed the so-called effect of
self-confinement of CRs. CRs excite turbulence by
the instability (\ref{stinst}) to the level when the streaming
velocity of CRs equals the velocity of self-excited waves.
CRs are frozen into these fluctuations and move with their
velocity.

In Breitschwerdt et al. (1991, 1993), and later in Zirakashvili et
al. (1996), the system of cosmic ray hydrodynamic equations was analyzed,
describing the balance of mass, momentum and energy together with
the energy balance of CRs and hydromagnetic fluctuations in the Galaxy.
They showed that, indeed,
the convective velocity is small near the galactic plane, but the
wind is accelerated by gaseous, CRs and MHD wave pressures,
and as a result the wind velocity reaches the value of several
hundred km s$^{-1}$ far away from the Galactic plane. Numerical
calculations showed that the wind velocity increased almost
linearly with distance close to the Galactic plane.

If CRs couple to the plasma via waves, a galactic wind develops,
and CRs are picked up at the height $z_c\sim D/V$ by the wind
and carried away from the galaxy. Hence the \emph{convective boundary}
is of the same importance as a \emph{run-away boundary} for free
particle escape. In general, $z_c$ is also a function of energy.

However, attempts to analyze the CR chemical composition
in the framework of an accelerating wind resulted in an unexpected conclusion.
According to the wind model, the convective boundary $z_c$
should be fairly close to the Galactic plane, i.e. $z_c\leq 1$ kpc
(see Bloemen et al. 1990). The problem was that in this case, it was impossible
to reconcile the chemical composition data with the almost unform CR
distribution in the galactic plane, inferred from EGRET
observations (see Strong and Mattox, 1996). The reason is the following:
the spatial distribution of CR sources (SN remnants, superbubbles) is
strongly nonuniform in the disk. It has a peak at a Galactocentric radius
between 4 and 6 kpc (coinciding with the observed molecular ring of the Galaxy),
and drops off exponentially at larger distances.
If the convective boundary is very close to the
Galactic plane, the CR density is determined by the
local source density, and therefore their distribution in the
disk should follow closely the source distribution.

The only viable solution in the framework of the diffusion model
is an efficient mixing of particles from different sources,
implying a huge CR diffusion halo in order to wipe out the
observed $\gamma$-ray gradient. However, even in the most
favorable case of an extremely extended halo, the standard model
is unable to remove the signature of the source distribution (see
Dogiel and Uryson 1988,  Bloemen et al. 1993). 

One way to escape this dilemma is to assume that some of the observational data are
not significant, like e.g. the SN distribution derived from radio
observations because of absorption (Strong et al. 2000). However,
Dragicevich et al. (1999) and Sasaki et al. (2004), analyzing SN
distributions in external galaxies (where the effect of absorption
is negligible), found similar radial SNR distributions as in our Galaxy.
Therefore, the alternative is to conclude that it is time to
abandon the standard diffusion model.

Paradoxically, as it may seem, the reason for a uniform CR
distribution may indeed be a convective boundary close to the
disk, if one accepts the possibility that CRs are also transported
into the halo locally as a result of a chimney type flow. In,
e.g., Bloemen et al. (1993) it was assumed that the location of
the boundary is independent of Galactocentric radius.  However,
from a physical point of view, it is more reasonable to assume
that the boundary is determined by the pressure driving the
outflow, i.e., it the higher the pressure the closer it is to the
disk. This was the ansatz of the model of Breitschwerdt et al.
(2002), who were able to show, that a higher SN rate led to a
rapid transport of CRs out of the Galaxy and thus reduced the
$\gamma$-ray flux locally. In their comprehensive analysis of 1-D,
2-D and 3-D convection models they were able to explain an overall
shallow $\gamma$-ray gradient as a result of a quasi-uniform
distribution of CRs in the galactic disk.

The main effect can be described in the following way. The
pressure of CRs is proportional to the density of CR sources,
$P_{cr}\propto Q$. Since the convection velocity $V$ is
proportional to the pressure, $V\propto P_{cr}\propto Q$, CRs
leave the Galactic disk faster from a region of higher source
density, and therefore the lifetime of CRs there is shorter than
in other parts of the disk. The density of CRs in the disk can be
estimated as $N_{cr}(r)\propto Q(r)/T_{cr}(r)$. We see that a high
rate of CR production by source at any galactocentric radius is
compensated by faster CR remove from this region. This effect
cannot be obtained in the framework of a standard diffusion model.
In this respect the "CR gradient in the disk" is an important
piece of evidence in favor of a galactic wind in the Galaxy.
Recently, Everett et al. (2008) have presented further support for
a CR driven galactic wind by showing that soft X-ray observations
in a certain Galactic region can be better explained by a wind
than by a hydrostatic halo model.

In a companion paper (Breitschwerdt et al. 2008, these proceedings) we will discuss
the properties of galactic winds, and show that soft X-ray halos in external galaxies
can be naturally explained in a self-consistent galactic wind model with non-equilibrium
ionization structure.

\begin{acknowledgements}
VAD  is partly supported by the RFBR grant 08-02-00170-a, the
NSC-RFBR Joint Research Project No 95WFA0700088 and by the grant
of a President of the Russian Federation ``Scientific School of
Academician V.L.Ginzburg''. DB thanks the organizer for some financial support.
\end{acknowledgements}

\end{document}